\documentstyle[preprint,aps,epsf]{revtex}
\begin{document}

\title{THEORY OF COEXISTENCE OF SUPERCONDUCTIVITY AND FERROELECTRICITY:
A DYNAMICAL SYMMETRY MODEL}

\author{
{\bf Joseph L. Birman} \\
Physics Department, The City College, CUNY,\\
Convent Ave. at 138 St, New York, NY 10031, USA\\
{\bf Meir Weger}\\
The Racah Institute of Physics,\\ The Hebrew University
Jerusalem ,Israel} 
\date{\today}
\maketitle

\begin{abstract}

We propose and investigate  a model for the  coexistence of 
Superconductivity 
(SC) and Ferroelectricity (FE) based on the dynamical symmetries
$su(2)$  for the (pseudo-spin) SC sector , $h(4)$   for the 
(displaced oscillator) FE 
sector and $su(2) \otimes h(4)$ for the composite system. We assume a 
minimal  symmetry-allowed coupling ,and simplify 
the hamiltonian using a  double mean-field approximation (DMFA). A  
variational coherent-state (VCS) trial wave-function is used for the 
ground state : the energy, and  the relevant order parameters for SC 
and FE are obtained . For positive sign of the SC-FE coupling 
coefficient,  a  non-zero value of 
either order parameter can suppress the other one (FE 
polarization suppresses SC , and vice versa). This gives some support to   
"Matthias' Conjecture" [1964] , that SC and FE tend to be mutually 
exclusive. For such  a  Ferroelectric 
Superconductor we predict: a) the SC gap $\Delta$ (and $T_c$)
will increase with increasing applied pressure when pressure quenches FE,
as in many ferroelectrics  and  b) 
the FE polarization $\vert \vec{P} \vert $ will increase with increasing 
applied magnetic field up to $H_c$, which  is equivalent to the 
prediction of a new type of Magneto-Electric Effect in a coexistent SC-FE
material.  Some discussion will be given of possible relation of these 
results to  the cuprate superconductors . 


\end{abstract}
\newpage
{\bf I.Introduction, and Background.}

This paper concerns an investigation of  coexistence of Superconductivity 
(SC) and lattice Ferroelectricity (FE) based on a model which 
expresses the Dynamical Symmetry underlying the physics.
The model combines the dynamical symmetry of the SC and FE  sub-systems 
into that for the   composite system. 

\parindent 3em

In Section II we review the algebra of the "s-wave" pairing BCS model 
of a  superconductor . It is well  known that a dynamical symmetry $su(2)$ 
algebra can be identified based  on time-reversed electron pair operators . 
The mean-field reduced Hamiltonian will be an element in this algebra. By 
rotating the  Hamiltonian in the space of the generators to a "diagonal " 
form, the   energies , the eigenfunctions and  then the  expectation 
value of the SC  order parameter in the ground coherent state, is obtained. 

\parindent 3em

In Section III we introduce a simplified algebraic model which has 
$h(4)$dynamical symmetry  for a displacive ferroelectric.
It is a "displaced oscillator" model for the phonon soft-mode which has 
represening the soft transverse-optic (TO) 
mode. This Hamiltonian can also be transformed to "diagonal" form
to give the energies,the eigenfunctions, and  the FE order parameter in 
its ground coherent state. 

\parindent 3em

In Section IV we introduce the SC-FE coupling ,and we discuss interactions 
which  will respect the gauge and inversion symmetries that will be 
broken.The Hamiltonian of the composite system including 
initially biquadratic interaction, is simplified using a  "double mean 
field  approximation ". The resulting bilinear Hamiltonian 
$\hat{H}_{DMFA}$ is in the direct product $su(2) \otimes h(4)$ algebra, and 
appears to be  the simplest way  that the two sub-symmetries can be joined.

\parindent 3em

In Section V we show that although  this Hamiltonian cannot be solved 
exactly, a variational solution can be found, by forming a trial  
eigenfunction analogous to the product coherent state. After carrying out 
the variational solution, the energy  spectrum, and the eigenfunctions 
are obtained. 

\parindent 3em

In Section VI we calculate the 
expectation values of the SC and the FE order parameters in the ground 
state of the coupled system. Our results for the order parameters show  
that : the presence of one non-zero
order parameter (e.g. spontaneous polarization) tends to suppress the 
other (e.g. superconductivity) and vice-versa in the  case of positive 
sign (repulsive) of the  coupling between the two  
subsystems. This leads to the prediction that the superconducting 
critical temperature can increase with pressure if FE is quenched by 
pressure , and is in agreement with experiment in sodium tungsten bronze. 
Another prediction is that  the ferroelectric polarization will increase 
with applied magnetic field . This is a new type of Magnetoelectric Effect.
These predictions are discussed in Section VI.

\parindent 3em

In the final Section VII we discuss the results especially for the sodium
tungsten bronze, and doped $SrTiO_3$ systems and we  suggest the  
relation of the present  work to the mechanism of high temperature 
superconductivity in 
the copper-oxide systems as previously proposed  by Peter and Weger  
\cite{PE1},\cite{PE2} based on the close proximity of  a 
near-ferroelectric instability to the superconducting transition. 
\parindent 3em

We believe the model presented here offers a simple "generic" way to 
treat the two broken symmetries relevant to the problem: for SC, broken 
gauge  symmetry ( the basic pair operators do not conserve number of 
electrons), for FE broken inversion symmetry (spontaneous TO phonon 
displacement is not $i$-invariant). In passing we will 
show  later that , if certain couplings vanish ,  our model can reduce  to 
other, well-studied models:  the Jaynes-Cummings Model , and the 
Spin-Phonon  model. Our model is  more general than either of these two, 
and truncating our  model to obtain  either of them would result in 
losing relevant physics for our case. 

\parindent 3em

Apart from interest  in new models for competing phases ,  there are 
several   reasons 
for  work  on  this problem at this time. As far back as 1964 , Matthias and 
coworkers 
\cite{MA1} reported superconductivity with $T_c$ below $1^o K$ in a sodium 
tungsten bronze  $Na_x WO_3$ with $ 0.1 \leq x \leq 1$.
These authors remarked, the host crystals are isomorphic (at $ x=1$ ) 
to barium titanate,so that it is  "probable that they 
are also Ferroelectric,in the sense of developing a polar axis, similar 
again to to $BaTiO_3$ and $WO_3$" . Matthias (1949,1967) and Matthias and 
Wood (1951) \cite{MA1} had confirmed the polar state for the sodium tungsten 
bronzes which they studied. Subsequent work of S.C.Abrahams et. al. 
\cite{AB1}, and others,  has also reported  on the  structural phase 
transitions and the development 
of a polar axis in many of these compounds. The doped tungsten bronzes 
have also  been studied as examples of the Mott and metal-insulator 
transitions \cite{MO1}. In  undoped  $WO_3$ , five phase transitions have 
been identified: at 40, 65, 130, 220, and 260 $ ^oK$ 
\cite{LE1}, \cite {HO1}, and most recently such transitions were studied 
by by Aird et.al.\cite{AI1}. Several recent reports of high temperature 
superconductivity
$T_{c} \sim 90 ^{o} K$ in a sodium tungsten bronze $Na_{0.05}WO_{3}$
system \cite{RE1}
added  considerable new stimulous for this work,  since we can suppose that 
the system is in a polar state at the superconducting temperature , which 
allows ferroelectricity.

\parindent 3em

It is worth recalling, too,  that SC-FE coexistence, and competition
played a  role in motivating the work of Bednorz and Muller  \cite{BE1} on
the high temperature cuprate superconductors.  Even earlier, work on the
"old " superconductors of the $\beta -W$ structure like $V_3 Si,Nb_3 
Sn$,etc., where $T_c \sim 23 ^oK$, and a martensitic phase transition occurs in
the same temperature range  ,  gave rise to investigations on the
possibility of a "ferroelectric metal"  or a "polar metal" , and thus to
the  study of SC-FE coexistence 
\cite{MA5},\cite{AN1},\cite{BI1},\cite{TI1},\cite{BHA1}. 

\parindent 3em

A number of theoretical papers have already discussed microscopic  models 
for  the  effect of lattice instability on superconductivity in the sodium 
tungsten  bronze systems \cite{NG1},\cite{NG2},\cite{KI} .
These papers have illuminated many  aspects of the interplay between 
the  structural deformations , such as rotation of underlying octahedaral 
units,  and coupling with electron pairs. The present work 
looks at the same problem of coexistence of SC and FE from a dynamical 
algebra , or  "Spectrum Generating Algebra [SGA]",  point of view , which 
complements these detailed models .

\parindent 3em

Study of  SC-FE coexistence problems  can be 
relevant to  recent work by Weger  and collaborators 
\cite{PE1},\cite{PE2},  on 
the mechanism of high temperature superconductivity in the
cuprates. In  that work the presence of a nearby FE instability close to 
the SC transition is related to the anomalously large  ionic 
dielectric coefficient in the cuprates, which reduces the electron-electron 
repulsion and can then  lead to an enhanced net electron-electron 
attraction, producing higher $T_c$.

\parindent 3em

The present  work  also relates   to earlier dynamical symmetry 
investigations by Birman and Solomon 
\cite{BI2},\cite{BI3},\cite{BI4},\cite{BI5},\cite{AS1},\cite{AS2},
\cite{AS3},\cite{AS4},\cite{AS5} 
for systems with multicritical 
behavior involving superconductivity and charge and spin density waves. 
Based on the mean-field models which had earlier 
been used to investigate superonductor-charge density wave , and 
superconductor- antiferromagnetic coexistence, these papers  were 
the   first  to introduce  models with $S0(6)-S0(5)$ symmetry for 
Superconductor plus Charge Density Waves (SC-CDW), and Superconductor plus
Antiferromagnetism 
(SC-AFM), as well as presenting and analysing
a general $SU(8)$ "Grand Unified Theory (GUT)"  model unifying singlet 
and triplet Superconductivity, and Charge and Spin Density Wave  
cooperative effects \cite{AS4}.  
Recently , SO(5) and SU(4) , models for multicritical  
superconductor-antiferromagnetic behavior in the high  temperature 
superconductors 
have also   been studied by Zhang, Demler, Guidry , and others 
\cite{SZH1},\cite{SZH2}.

\newpage
{\bf II. The $su(2)$ Pseudospin  Model For  a Superconductor.}

The BCS theory for superconductors  can be conveniently epitomized at 
a mean-field level  by introducing the pseudo-spin  $su(2)$ 
algebra of the fermion pair operators 
\cite{BA1},\cite{PWA1},\cite{TIN},\cite{W1}. Since this is 
well known, here we briefly summarize the material needed for later 
reference. 
Some more notational and other details are given in  Appendix 3.  

\parindent 3em
In the dynamical symmetry $su(2)$ model for a superconductor  we
take the Hamiltonian in a reduced mean field approximation as 

\begin{equation}
\hat{H}_{SC} \  = \  \sum_k \hat{h}_k
\end{equation}

\noindent
with the hamiltonian at sector k given as
\begin{equation}
\hat{h}_k \ = \ -2\epsilon_k \hat{j}_{3k} + 2\Delta_k\hat{j}_{2k} 
+2\epsilon_k
\end{equation}

\noindent
Here $\epsilon_{k}$ is the single electron energy, with $\epsilon_{k}$ \
= \ $\epsilon_{k\uparrow}$ \ = \ $\epsilon_{-k\downarrow}$, and 
$\Delta_{k}$ is the  pairing ("gap") energy . The dynamical symmetry , or 
spectrum generating 
algebra (SGA) , for each $\hat{h}_{k}$ is $su(2)_{k}$ , so 
the SGA  of 
the entire Hamiltonian is $\bigotimes_{k} su(2)_{k}$ .When it causes no 
confusion we drop the index $k$.   

\parindent 3em
 	
	The $su(2)$  pseudo-spin operators $\hat{j}_{pk}$ obey

\begin{equation}
[\hat{j}_{pk},\hat{j}_{qk}]_- \ = \ i \epsilon_{pqr}\hat{j}_{rk}, \ {\rm 
where}
   \ \ (p,q,r) = (1,2,3)
\end{equation}

\noindent
and each $\hat{j}_{pk}$ is a bilinear in the fermion operators ( see 
Appendix 3). When all 
$\Delta_{k} =0$, each $\hat{h}_{k}$ is the hamiltonian for free electrons
and the total wave function in the ground state is simply the product of 
individual creation operators acting on the vacuum state for that $k$. 
The filled Fermi Sea is the many-electron ground state , or the 
"disordered" state , denoted $\vert 0 \rangle = \prod_{k} 
{\hat{a}\dagger}_{k{\uparrow}}{\hat{a}\dagger}_{-k{\downarrow}} \vert 0 
{\rangle}_{k}$. For $\Delta_{k} \not= 0$ the ground state is obtained at 
each k, by rotating $\hat {h}_{k}$ about the $\hat{j}_{1k}$ axis by the 
angle $\theta _{1k} = tan^{-1}({\Delta_{k}/ \epsilon_{k}})$ , so that the 
transformed hamiltonian will be parallel to the $\hat{j}_{3k}$ axis.

	 The rotation operator is given by

\begin{equation}
\hat{U}_{1k} \ \equiv \ exp(i\theta_{1k}\hat{j}_{1k})
\end{equation}

\noindent
so that

\begin{equation}
\hat{h}^{\prime}_{k} \ = \ \hat{U}_{1}\hat{h}_{k}\hat{U}^{-1}_{1} \ = \
E_{k}\hat{j}_{3k}.
\end{equation}

\noindent
Now $\hat{h}^{\prime}_{k}$ is directed along $\hat{j}_{3k}$ and its
eigenvalue is

\begin{equation}
E_{k} \ = \ \sqrt{\Delta^{2}_{k} + \epsilon^{2}_{k}}.
\end{equation}

\parindent 3em
	The key ingredient needed in order to discuss a phase 
transition is the order operator $\hat{\eta}_{SC}$ for superconductivity.
A natural choice is the real part of the pair operator, i.e. 
$\hat{\eta}_{SC} =  \hat{j}_{2k}$.Then, the order parameter in a state is 
the expectation  value of that order operator in that state. We  
single out two states : the disordered state $\vert \Psi _{dis} \rangle$
=$\vert 0 \rangle$, 
and the ordered state $\vert \Psi 
_{ord} \rangle $. Using the identifications given above for each of these 
states  , depending on whether $\Delta _{k} = 0 ,or \not= 0$ , we have
$\eta_{SC} =0 $ ,or $\eta_{SC} \not= 0$ respectively.We can go further . 
When $\Delta_{k} \not= 0$, the ordered state is $\vert \Psi_{ord} \rangle =
\prod_{k} \vert \psi_{ord,k} \rangle = 
\prod_{k} \hat{U}^{-1}_{1k} \vert jm \rangle $ where $\vert jm \rangle $
is the ususal eigenstate of $su(2)$ (for details  see Appendix 3). Then, in 
the ordered state we have
	
\begin{eqnarray}
\eta_{SC} & \equiv &
\langle\psi_{k}\vert \hat{j}_{2k}\vert\psi_{k}\rangle  \\
 & = &  \langle jm\vert\hat{U}_{1}\hat{j}_{2k}\hat{U}^{-1}_{1}\vert
jm\rangle  \\
 & = &  \langle jm\vert(\hat{j}_{2k} \cos\theta_{1} +
\hat{j}_{3k}
\sin\theta_{1})\vert jm\rangle \\
 &  = &   m \sin\theta_{1}
\end{eqnarray}

\noindent
or
\begin{equation}
\eta_{SC}  =  m \Delta_k/\sqrt{\Delta^{2}_{k} +
\epsilon^{2}_{k} }.
\end {equation}

\noindent
The lowest (ground state) energy  occurs when
$ m \ = \ -1/2$, and the  order parameter $\vert\eta_{SC}\vert
 \not = 0 $ in this state.

\parindent 3em

        The states $\vert{\psi_{k}}\rangle
= \hat{U}^{-1}_{1}\vert {jm}\rangle $, and in particular the ground state
$ \vert{\psi_{kG}}\rangle $ are $su(2)$ coherent states in accord with  the
usual  definitions 
\cite{AP1},\cite{AS6},\cite{RA1},\cite{KL1}. 
Globally the symmetry of the total hamiltonian
$\hat{H}$, which is $\otimes_{k} su(2)_{k}$ , gives a total wave
function of the ground state which is the global coherent state
\begin{equation}
\vert{\Psi_{BCS}}\rangle    =
\prod_{k} \vert {\psi_{kG}}\rangle
\end{equation} .

\parindent 3em

	The steps used in this Section will be used again below. Namely,
we: a) identify the dynamical symmetry of the hamiltonian chosen; b)
diagonalize the hamiltonian ,thus obtaining the ground and excited state 
eigenfunctions (coherent states) and energies ; c) evaluate the 
expectation of the  relevant  order operator in the appropriate state to 
give the order parameter.

\newpage 
{\bf III.Algebra h(4) for a Displacive Ferroelectric Soft Mode.}
	
Ferroelectricity in perovskite-like systems is due to a "soft phonon"
transverse optic (TO) lattice mode resulting from ionic displacements that 
break inversion symmetry \cite{CO1},\cite{PWA2},\cite{COW1}, \cite{LI1}. 
When these displacements 
are "frozen" by higher order anharmonic terms  stabilizing a 
distorted structure, a macroscopic FE polarization $\vec{P}$ arises; the 
magnitude  $\vert{\vec{P}}\vert$ is proportional to the expectation 
value of 
the frozen soft mode amplitude $\langle {\vec{\hat{Q}_{0}}} \rangle$. Here 
$\vec {\hat {Q}_{0}}$  
is the operator of the normal coordinate displacement for the soft mode 
\cite{BH1}. In terms of the harmonic oscillator boson operators 
$\hat{B}_{0}$,
$\hat{B}^{\dag}_{0}$ of the soft mode, $ \vec{\hat{Q}_{0}} \sim \ 
({\hat{B}}^{\dag}_{0} +
{\hat{B}}_{0} ) $. We will assume  that only a small 
region in the $ \vec{k}$-space softens, and we can take that to be near 
$\vec{k} = 0$ . Thus we consider a homogeneous ferroelectric material 
, and we neglect FE density waves or stripes which may occur if $ k \not= 
0$. The soft mode will be represented by a harmonic oscillator.

\parindent 3em

In the presence of spontaneous polarization $\vec{P}$, a 
macroscopic self-electric field $\vec {E}$ arises , giving an 
energy  proportional to $\vec E \cdot \vec P$ \cite{LA1}. The field $\vec 
{E}$ is the internal transverse electric field generated by the 
transverse polarization . The origin of the latter is the frozen 
$TO$ mode. Translating 
this  self-term into the language of our problem we shall write this energy 
term as $\gamma_{1} {\cal{E}}(\hat{B}^{\dag}_{0} + \hat{B}_{0})$ , where
$\gamma_{1}$ is a coupling constant we shall take as positive.

\parindent 3em

We  then take as our model for 
the ferroelectric sector of the hamiltonian  the sum of the harmonic 
oscillator term for the soft mode ,  plus the energy due to the 
self-field coupled to
the polarization ; higher order anharmonic terms are omitted. In second 
quantized form this Hamiltonian is :

\begin{equation}
\hat{H}_{FE}  = \omega_{T0}(\hat{B}^{\dag}_{0} \hat{B}_{0} + 1/2) +
\gamma_{1} {\cal {E}} (\hat{B}^{\dag}_0 + \hat{B}_0).
\end{equation}
The frequency of the soft TO mode is $\omega_{T0}$, we take  $\gamma_1 $ 
as a 
positive constant , and ${\cal {E}}$ is the magnitude of $ \vec E$.  
We immediately recognize $\hat H_{FE}$ as  a"displaced oscillator" 
hamiltonian for the soft mode, including coupling to the self-macroscopic 
field. 

\parindent 3em

Before proceeding, note that the theory of ferroelectricity for
soft  mode perovskite-like systems has a long history. Microscopic models
including anharmonic terms in the Hamiltonian  followed the original
work of Cochran (1959) \cite{CO1} and Anderson \cite{PWA2}  who proposed the 
"soft mode" model . The models were carefully analysed
by Cochran \cite{CO2}, Cowley \cite{COW1}, Cowley and Bruce \cite{COW2}, 
 and others. Soon after  experimental work on the doped tungsten 
bronzes was reported , several detailed theoretical papers appeared giving 
various  microscopic models for the interplay between SC and FE , and 
also related  to the Metal-Insulator transition in these materials 
\cite{MO1},\cite{HO1}, \cite{SH1},\cite{NG1},
\cite{NG2},\cite{KI} , and active study  continues to the present 
\cite{NS1}.

\parindent 3em

Now we  briefly comment  about the soft mode. In the 
context of a traditional soft-mode displacive ferroelectric such as
a perovskite like $BaTiO_3$ or a tungsten bronze like $A_{x}WO_3$ the soft 
mode can exhibit a typical temperature dependence such as:

\begin{equation}
\omega^{2}_{T0}  =    \Omega \vert (T-T_{c}) \vert 
\end{equation}

\noindent
with $\Omega $  a constant , and $T_{c}$ the FE transition 
temperature. On the other hand , in their very recent 
investigations on the mechanism of High Temperature Superconductivity in 
the cuprates, Weger and  collaborators \cite{PE1},\cite{PE2}, have examined 
quantitatively how the electron-phonon and electron-electron interactions
are affected by the medium itself having a high lattice-induced dielectric
coefficient. For example , values of $ \epsilon (\omega)  \approx  50 
$ or more  for $\omega = 10 mev $ , in 
$La_{2-x} Sr_{x} Cu O_4$ and YBCO were recently measured 
\cite{KI1},\cite{HE1}.
As noted by these authors,  the relevant soft-mode 
transverse  optic (TO) phonon is  governed by the 
Lyddane-Sachs-Teller (LST) relation \cite{BH1} which for a single pair 
of LO and TO modes is:
	
\begin{equation}
\epsilon ( \omega)  = \epsilon(\infty) ( \omega^2 - \omega^2_{LO}) / 
(\omega^2 - \omega^2_{TO})
\end{equation}

Here $\epsilon(\infty) $ is the high frequency dielectric coefficient, and 
$\omega_{LO,TO}$ are the characteristic phonon frequencies. 
This single mode expression  works well for the c-axis component of  the 
modes in both $ La_{2-x}Sr_{x}CuO_4$ and YBCO .
In  our work   we will not use the explicit expression given in  equation 
(14) , but  we shall make  contact with the idea  that the dielectric 
coefficient at low frequencies  is very large , which is
related to the  "near ferroelectric instability".

\parindent 3em

We return to the Hamiltonian for the ferroelectric sector, $\hat{H}_{FE}$
It is well known \cite{KL1},\cite{NI1}, \cite{GL1}, that the "displaced 
oscillator hamiltonian" can be transformed  by the unitary operator

\begin{equation}
\hat{U}_{2} \ = \ exp [\xi_{0}(\hat{B}^{\dag}_{0}-\hat{B}_{0})].
\end{equation}

\noindent
which displaces the oscillator Bose operators as

\begin{equation}
\hat{U}_{2}\hat{B}^{\dag}_{0}\hat{U}^{-1}_{2} = \hat{B}^{\dag}_{0} + 
\xi_{0}. \end{equation}

\noindent
We take the simplest case with $\xi_{0}$ real ,  and by choosing $\xi_{0}  = 
( -\gamma_{1}/\omega_{T0}) $ , we obtain the transformed hamiltonian as

\begin{eqnarray}
\hat{H}^{\prime}_{FE} & = & \hat{U}_{2}\hat{H}_{FE}\hat{U}^{-1}_{2} \\
& = &  [\omega_{T0}(\hat{B}^{\dag}_{0} \hat{B}_{0}+1/2) 
 -(\gamma_{1}{\cal{E}})^{2}/\omega_{T0}].
\end{eqnarray}

\noindent
This transformed oscillator hamiltonian $\hat{H}^{\prime}_{FE}$ is shifted 
to a new minimum, but 
retains the same excitation frequency $\omega _{T0}$ as the original 
oscillator.

\parindent 3em

We seek the eigenstates $\vert \Phi \rangle $ of the original hamiltonian

\begin{equation}
\hat{H}_{FE} \vert\Phi\rangle  =  W\vert \Phi\rangle.
\end{equation}

\noindent
Transforming this equation by the operator $\hat{U}_{2}$ and continuing 
as in Section II ,  we find the eigenvalues and eigenfunctions as

\begin{equation}
W_{n}  =  (n+1/2) - (\gamma_{1}{\cal{E}})^{2}/\omega_{T0}.
\end{equation}
and
\begin{equation} 
\vert \Phi \rangle  =  \hat{U}^{-1}_{2}\vert n \rangle   = 
exp-[\xi_{0}(\hat{B}^{\dag}_{0} - \hat{B}_{0})] \vert n \rangle .
\end{equation}
where $\vert n\rangle $ is a number eigenstate of the phonon number operator
$\hat{N_{B}}  \equiv  \hat{B}^{\dag}_{0}\hat{B}_{0}$. The state $\vert 
\Phi \rangle$ is a Glauber coherent state for the FE oscillator 
\cite{KL1},\cite{GL1}.

\parindent 3em

A natural choice of the order operator for the FE polarization is the 
coordinate operator $\hat{Q}_{0}$ , or $ (\hat{B}^{\dag}_{0} + \hat{B}_{0})$ 
. Thus $ \hat{\eta}_{FE} \sim  \hat{Q}_{0} \sim (\hat{B}^{\dag}_{0} + 
\hat{B}_{0})$.   Clearly, in 
the state $\vert n\rangle$ i.e. ,  $\vert \Psi_{dis} \rangle$ , of the free 
phonon the expectation value of this order operator is zero:
$\langle n \vert \hat{Q}_{0} \vert n \rangle $= 0 , meaning no 
spontaneous polarization . But in the Glauber 
coherent state $\vert \Phi\rangle $ , which is $\vert  \Psi_{ord} \rangle 
$we have a 
non-zero value of the order parameter as the following argument shows:

\begin{equation}
\eta_{FE}  =  \langle  \Phi \vert \hat{Q}_{0} \vert \Phi \rangle  = 
\langle n \vert \hat{U}^{-1}_{2} (\hat{B}^{\dag}_{0} + \hat{B}_{0}) 
\hat{U}_{2} \vert n \rangle  =  2\xi_{0}  =  -2(\gamma_{1} {\cal{E}}  )
/\omega_{T0}. 
\end{equation}

\parindent 3em 

 If, further, we identify the magnitude of the order parameter with 
the macroscopic  polarization we have

\begin{equation}
\vert \eta_{FE} \vert  =  \vert \vec P\vert  =  (2\gamma_{1}{\cal{E}}) /
\omega_{T0}.
\end{equation}

\noindent
Although the field $\bf{E}$ is a self-field due to the spontaneous 
polarization, we may treat this expression as defining a macroscopic
dielectric susceptibility ,  given as $ \chi  =  (\vert \vec P 
\vert/{\cal{E}})  = 2 (\gamma_{1} / \omega_{T0}).$ For a soft mode with 
$\omega_{T0} \rightarrow 0 $ (for example as $ T \rightarrow T_{c})$
the susceptibility $\chi$ will become very large . This interpretation 
agrees with what one would expect of a 
ferroelectric, or "near-ferroelectric" transition. Higher order terms will 
stabilize the system and prevent actual divergence.

\parindent 3em
	
While the soft-mode displaced oscillator Hamiltonian is a very 
simplified version of the true state of affairs,
this hamiltonian $\hat{H}_{FE}$ captures the  physics of the FE 
sector for our purposes in this paper. Namely, this hamiltonian exhibits 
the FE  displacement of the oscillator ,which breaks a pre-existing 
inversion symmetry, and gives  the enhanced dielectric susceptibility, in a 
simple algebraic $h(4)$ setting. To include higher order anharmonic or 
coupled terms could make the model more realistic, but would depart from 
our algebraic framework.

\parindent 3em

Just as we do not expect the $su(2)$ SC model to be a microscopic 
model which can give all the features of 
superconductivity , so the $h(4)$ FE model does not claim
to be a microscopic model incorporating  interactions needed to explain all
properties of ferroelectric media.
\newpage
{\bf IV. Interaction Terms; Double Mean Field Approximation.}

\parindent 3em

{\bf a) Interaction Terms}

\parindent 3em

To procede we need the total Hamiltonian which we take as

\begin{equation}
\hat{H} = \hat{H}_{SC} + \hat{H}_{FE} +  \hat{H}_{INT}.
\end{equation}

We will take $\hat{H}_{SC}$ and $\hat{H}_{FE}$ as before, and now we turn 
to $\hat{H}_{INT}$ . For  the Hamiltonian to be translation invariant 
\cite{BI6}
the interaction term must match the wave vectors of the soft-mode phonon 
and the  Cooper pairs. Since the center of mass momentum of
the pairs vanishes, we will directly couple the $k=0$ lattice soft mode 
and the electron pairs. 

\parindent 3em

In order to find the form of $\hat{H}_{INT}$ in our algebraic 
framework, we need to use the basic operators in the SC and FE sectors, and
combine them in an invariant fashion. It is natural to be guided by 
general prescriptions used in the Landau approach \cite{LA2} , and 
especially in
the Ginzburg-Landau theory for competing order parameters. Here, for 
competing ferroelectricty and superconductivity,  
the order parameters are the spontaneous
FE polarization $\vert \vec {P} \vert$ and the superconducting gap
$\Delta$ respectively.  It is required , in the G-L theory ,that
every term in the Free Energy shall be a scalar invariant under the 
relevant symmetry group, which in the present work is the direct product 
of the configuration space
symmetry group of the crystalline medium, and the gauge group of the unbroken
many-electron sector. We take the  TO phonon in a prototype 
paraelectric crystal coupled to an "s-wave" SC complex gap parameter. For 
example take the prototype systems to be $ W0_{3}$ or perovskite-like , 
with cubic symmetry in the paraelectric and non-superconducting phase. The 
TO phonon will be split off from a $\Gamma(15^{-})$ or other 
three-dimensional
representation\cite{BI6}. In the  homogeneous  approximation where the 
order parameters are uniform , the Free Energy density of the system will 
be of the form

\begin{equation}
\delta F = a \vert\vec{P}\vert^{2} + (b/4) \vert \vec{P} \vert^{4} +
\alpha \vert \Delta \vert^{2} + \beta/4 \vert \Delta \vert^{4} +
\kappa/2 \vert \Delta \vert^{2} \vert \vec{P} \vert^{2}
\end{equation}

In such a  G-L theory , the lowest order, "generic" coupling  between 
superconducting  $\Delta$ and the ferroelectric $\vec P$  will be a 
biquadratic  term of the form proportional to $(\vert \Delta \vert^{2} 
\vec {P}^{2}) $, i.e. the last term in $\delta F$ above. This is the 
lowest order term which  satisfies gauge and 
parity symmetry requirements. A similar  G-L theory was analysed by Liu and 
Fisher \cite {LIU1}, Imry \cite{IM1}, Yurkevich, Rolov and 
Stanley \cite{YU1}, and is discussed in texts such as that by 
Vonsovsky, Izyumov, and Kurmaev \cite{VON1}. 
 A rich phase diagram can result, depending on the relative magnitudes 
and signs of the coefficients of quadratic , quartic, and biquadratic 
terms .A case of particular interest to us,
will occur for "antagonistic" order parameters for which $\kappa$ is 
positive. It is then possible to have four distinct stable regions in the 
phase diagram , including one for which there is coexistence of non-zero 
$\Delta$ and non zero $P$. Although our Hamiltonian will not 
contain  terms of  the fourth degree like  $\vert P \vert ^{4}$ 
or $\vert \Delta \vert^{4}$, we will carry over the positive sign 
of the  coupling constant denoted $\gamma_2$ below, corresponding to  
$\kappa$.

\parindent 3em  

We need to  translate this coupling term into the operators of 
our model. Thus we  we  must add to the free hamiltonian $ \hat{H}_{SC} + 
\hat{H}_{FE}$ , the interaction term ,  in which we use the 
correspondences $ \vert \Delta \vert ^{2}
 \sim (\hat{j}_{2})^{2}$ , and $\vec{P}^{2} \sim \hat{Q}_{0}^{2} $  so that: 

\begin{equation}
\hat{H}_{INT}  =  \gamma_{2}(\hat{j}^{2}_{2})(\hat{Q}^{2})
 =  \gamma_{2}(\hat{j}^{2}_{2})(\hat{B}^{\dag}_{0} +\hat{B}_{0})^{2}.
\end{equation}

\noindent
We take $ \gamma_2 $, which corresponds to the $\kappa_2$,  to be a 
positive coupling constant, as 
discussed above , in order to implement the "antagonism" of the two 
ordering fields. In principle both  a  microscopic theory of 
the basic interaction, and comparison with experiment ( see below in 
Section VI)  will be needed in order to decide on the sign of $\gamma_2 $ 
for a particular system.

\parindent 3em

It  should also be noted that if the initial 
paraelectric phase is of lower symmetry than cubic  , a term of lower
degree for example ,  $\vert \Delta \vert^{2} P$ could occur. This would 
preserve gauge invariance , and use the component of $\hat P $ transforming 
like the identity representation of the paraelectric group.
Another manner in which a coupling linear in $\hat P $ might  arise,  uses
the fact that the soft mode  displacement, and the associated 
ferroelectric polarization $\hat P $ is transverse   in 
the "long-wave" limit \cite{BH1} . So, in order to couple to $\hat P $ we 
should seek a transverse field  operator associated to the 
superconductive sector. A natural candidate is the operator for the 
transverse superconducting current. Such a coupling term could have the form

\begin{equation}
\hat{H}^{'}_{INT}  \sim \hat{P}_{FE} \cdot\hat{J}_{SC}
\end{equation}

\noindent
where the SC current operator would have the form
\begin{equation}
\hat{J}_{SC} =(-ie^{*}{\hbar}/2m)[\psi^{*}\nabla \psi -\psi \nabla \psi^{*})-
(e{*}^{2}/mc)\vert\psi\vert^{2}\bf{A}
\end{equation}
\noindent
where $e^{*}$ is the SC pair charge and $\bf{A}$ is the vector potential.
In the London approximation $\bf{J}_{SC}$ $\sim$ $n_{SC}\bf{A}$ where 
$n_{SC}$ is the superconducting electron density , and $\bf\nabla \times 
\bf{A} = \bf{h}$ , with  $\bf{h}$  the local magnetic field.
Such a term which we can also write proportional to $\bf{P_{FE}}\cdot 
\bf{A}$ 
would  be symmetry allowed in the correct geometry (lower than cubic , and
with pre-existing broken ${i}$-symmetry). It would be linear in the 
transverse 
polarization. We  could rewrite this term to bring it to the form of the 
operators of our model if we further take ${J}_{SC}$ proportional to the 
number of pairs, as  $\Sigma \hat{b}^{\dag}_{k}\hat{b}_{k}$ and then 
express these in terms of the 
$\hat{j}_{1,2}$ of the SC sector, while $\hat{P}$ would be given in  
terms of the Bose creation and annihilation operators $\hat{B}^{\dag}_{0}$, 
and $\hat{B}_{0}$ respectively. However , since   these terms are 
linear in $\bf{P}$ and refer to pre-existent 
broken inversion symmetry , we do not include them here. 

\parindent 3em

 The  microscopic theory of the  electron-soft ferroelectric mode (TO 
phonon) interaction was studied by
Epifanov, Levanyuk and Levanyuk \cite{EP1} . They  point out that 
in the continuum
approximation there is no transverse optic phonon to  electron 
coupling because the TO mode does not induce a "macroscopic"
electric field. By  contrast, a longitudinal mode
  will give such a macroscopic field through
the effective charge:  $\nabla \cdot \hat{P} = -4\pi \rho_{b}$ where 
$\rho_{b}$ is an effective "bound" charge. But,  as pointed out in 
\cite{EP1},  using the correct lattice theory \cite{BH1},  there will  be 
coupling of the band electron  to the local internal electric field due to 
the TO displacement. Taking  the effective mass approximation 
\cite{BIP1} , a meaningful  continuum approximation can be , and was , 
defined  for the electron-soft TO mode interaction
 and is discussed in Epifanov et.al.
\cite{EP1}.

\parindent 3em

 In very recent work Fay and Weger \cite{WF1}  discuss  the  
renormalization of the electron-phonon vertex  in media  with large 
dielectric constant near the ferroelectric phase transition.

\parindent 3em

{\bf b) Double Mean Field Approximation}

\parindent 3em

Assume now that the system is in a state $\vert \psi \rangle$ such that  
the  product of 
the squares of the fluctuations of the operators $\hat{j}_{2}$ , and  $\hat 
{Q}$ evaluated in 
that state can be neglected. Thus if $\hat{A}$ and $\hat{B} $ are such 
operators and we write 

\begin{equation}
\hat{A}  =  [\hat{A} -\langle {A} \rangle] + \langle {A} \rangle 
 \equiv  \delta\hat{A} +\langle {A} \rangle
\end{equation}  
and similarly for $\hat {B}$, then neglecting $\langle( \delta \hat 
{A})^2 \rangle \times \langle (\delta \hat {B})^2 \rangle $ , we will have:

\begin{equation}
\hat{A}^2\hat{B}^2 \ \approx  (2\hat{A}\langle\hat{A}\rangle - \langle 
\hat{A} \rangle^2) (2\hat{B} \langle \hat{B} \rangle - \langle 
\hat{B}^2\rangle)
\end{equation}
Applying this to the biquadratic  interaction term we obtain

\begin{equation}
\hat{H}_{INT} \sim  \hat{j}^{2}_{2}(\hat{B}^{\dag}_{0} + \hat{B}_{0})^{2}  
\approx 4\Delta  P \hat{j}_{2}(\hat{B}^{\dag}_{0} + \hat{B}_{0}) -
2\Delta P^{2}\hat{j}_{2} - 2P\Delta^{2}(\hat{B}^{\dag}_{0} + \hat{B}_{0})
+ \Delta^{2} P^{2}
\end{equation}
Our  double mean field approximation (DMFA) yields a  bilinear effective 
interaction term $\hat{j}_{2}\hat{Q}$, and it  renormalizes the 
coefficients of the SC pairing term $\hat{j}_{2}$ and of the 
linear self-term in $\hat{Q}$, and in addition there is an  energy shift term
proportional to $\vert \Delta \vert^2 P^{2}$. In Appendix 1 we examine 
the validity of the DMFA using the variationally determined wave functions.

\parindent 3em

Recapitulating, we assume that initially we have the hamiltonian
\begin{equation}
\hat{H}= -2\sum_{k}(\hat{j}_{3k} + 2 \Delta_{k}\hat{j}_{2k}) +\omega_{0}
(\hat{N_B}+ 1/2) +  \gamma_{1}{\cal{E}} (\hat{B}^{\dag}_{0} + 
\hat{B}_{0}) + \sum_{k} 
\gamma_{2k}(\hat{j}_{2k}^{2})(\hat{B}_{0}^{\dag} + \hat{B}_{0})^{2}.
\end{equation}

\noindent
The bare coefficients $\Delta$, and $\gamma_{1}$,
 refer to the prototype superconductor, and the prototype 
ferro-electric , and $\gamma_{2}$ is the initial pair-TO mode coupling 
coefficient.  After making 
the double mean field approximation , and isolating a single mode $k$ ,  we 
have  the effective Hamiltonian at mode $k$ in the DMFA  :

\begin{equation}
\hat{H}_{DMFA} \ = \ -2\epsilon\hat{j}_3 + 2\Delta^{\prime}\hat{j}_2 +
\omega_0(\hat{N}_B +1/2)  + \Gamma_{1}(\hat{B}^{\dag}_0 + \hat{B}_0) +
\Gamma_2 \hat{j}_2 (\hat{B}^{\dag}_0  + \hat{B}_0) + \gamma_2 P^2 \Delta^2.
\end{equation}
The renormalized coefficients are $\Delta'$, and $\Gamma_{1,2}$. 
This will be our working Hamiltonian . It includes the SC and the FE 
prototype systems and their coupling via  the  
soft-mode oscillator coupled to  the  pseudo-spin pairing Hamiltonian.
Note that , technically , the initial hamiltonian is in the "enveloping 
algebra" of $su(2) \otimes h(4)$ because of the biquadratic terms , while 
$\hat{H}_{DMFA}$ is an element in the direct product algebra $su(2) 
\otimes h(4)$. Extensions of our model to include various higher order terms
coupling the phonon modes (anharmonicity) and pseudo-spin (pair-pair) terms
will be treated elsewhere. 

\parindent 3em
The values of the renormalized coefficients in $\hat{H}_{DMFA}$ are , in
terms of the original coefficients:

\begin{eqnarray}
\Delta^{\prime} & \equiv & \Delta(1-\gamma_2P^2) ; \nonumber \\  
\Gamma_1  & \equiv & \gamma_1 {\cal{E}} - 2(\gamma_2  P \Delta^2) ; 
\nonumber \\  \Gamma_2  & \equiv & 4 \gamma_2.
\end{eqnarray}

\noindent
When $\gamma_2 \rightarrow 0$ , we recover the sum of two 
separate  sectors: for SC and FE.

\parindent 3em
Our working model $\hat{H}_{DMFA}$ reduces  to 
two other well-known models in certain limits: the "Jaynes-Cummings" 
model \cite{SZ1}, and the "spin-phonon" model \cite{WEG1}, \cite{LEG1}. The 
single mode "spin-boson" or 
"spin-phonon" model  arises if $\Delta^{\prime}  = 0 $,    
and \ $  \Gamma_1 \ = 0$. Like our model, the spin-boson problem is not 
exactly soluble, but there has been much literature on it , including  very 
recent work. Another related model is 
the single mode  "Jaynes-Cummings" model
for a photon (boson) coupled to a two-level "pseudo-spin" atom. We obtain
the Jaynes -Cummings model by taking  $\Delta^{\prime}  =  0$ ,  and $  
\Gamma_1 \ = \ 0 $ , in $\hat H_{DMFA}$ , and  if we write 

\begin{equation}
\hat{j}_2 \ = (1/2)(\hat\sigma_{+} +\hat\sigma_-)  \ \equiv \ 
(1/2)(\hat{b}^{\dag} + \hat{b})
\end{equation}
where on the right hand side the $\hat{b}^{\dag}$ and $\hat{b}$ refer to 
the electron pair operators, 
and , in the bilinear coupling term $\hat{j}_2 (\hat{B}^{\dag}_{0}  \ +
\hat{B}_{0})$  and if we retain only the "energy-conserving" interaction 
, we obtain:

\begin{equation}
(\sigma_{+} \hat{B}_{0} + \sigma_{-} \hat{B}^{\dag}_0).
\end{equation}
This  is the rotating-wave approximation (RWA) to the Jaynes-Cummings model. 
However, we cannot make this approximation for the SC-FE problem, that is 
for our $\hat{H}_{DMFA}$, as we will lose   essential physics of 
our problem ; a similar point was  made in the review of Leggett 
et.al. \cite{LEG1} on the spin-phonon problem.

\newpage

{\bf V. The Variational "Coherent State" Eigenfunction}.

\parindent 3em

 In order to obtain the  
ground state eigenfunction and eigenvalue of our model $\hat H_{DMFA}$, 
we  will 
use a variational procedure based on the $su(2)\otimes h(4)$ symmetry.
Recall that when $\gamma_2 \ = \ 0$ the uncoupled Hamiltonian at mode $k$
is:

\begin{equation}
\hat{H} \ = \hat {h}_{SC} +\hat{H}_{FE}.
\end{equation}
Recall that this Hamiltonian is "diagonalized" by the product of two unitary 
transformations previously denoted 
$\hat{U} \ = \  \hat{U}_{1} \hat{U}_{2}$ where the parameters $\theta_1 = 
\tan^{-1}(\Delta)/\epsilon$ , and $\xi_0 = (-\gamma_{1}/\omega_{0})$ are 
the definite  values obtained  in Sections II and III . The 
resulting total state of the system without interaction   $\vert \Psi 
\rangle \ = \     
\hat{U}^{-1}_{1} \vert jm \rangle  \hat{U}^{-1}_2 \vert n \rangle $ is a 
coherent state which is the product of the coherent state of the pseudo-spin 
su(2) algebra for the SC sector, times the Glauber coherent state for 
the h(4) algebra for the FE sector.

\parindent 3em

For our  coupled $\hat H_{DMFA}$ at mode $k$  we introduce an  
analogous 
trial variational coherent state (VCS) which is the product of two 
"coherent-like"  states and is denoted $\vert \Psi_{v} \rangle $:

\begin{equation}
 \vert \Psi_v \rangle  \ = \ \hat{V} \vert  jm \rangle \vert n \rangle  
\ = \  \hat{V}^{-1}_{1} \vert  jm \rangle  \hat{V}^{-1}_{2}  \vert n \rangle
\end{equation}
where

\begin{equation}
\hat{V}_1 \  \equiv  \  \exp(i\theta \hat{j}_1 ) \ ; \hat{V}_2  \equiv \ 
\exp(\xi(\hat{B}^{\dag}_0 - \hat{B}_0))
\end{equation}
but now, the parameters $\theta $, and \ $\xi $ are variational unknowns.
The kets $\vert jm \rangle$ and $\vert n \rangle$ are as in Sections II and
III. We now define the energy in state $\vert \Psi_v  \rangle $  as the 
diagonal value of $\hat{H}_{DMFA}$ in the variational coherent 
state $\vert \Psi_{v} \rangle $

\begin{equation}
 \langle \Psi_v \vert \hat{H}_{DMFA} \vert \Psi_v \rangle  \ \equiv \ 
E_{m,n} (\theta , \xi )
\end{equation}
and we determine $\theta $ and $\xi $ from 

\begin{equation}
\partial E_{mn} /\partial \theta = 0  \ \rm and \ \ \partial 
E_{mn} / \partial \xi  = 0.
\end{equation}

The evaluation of $E_{mn}$ is particularly simple since the only 
non-zero
contributions to the diagonal value are from the matrix elements of the
operators $\hat{j}_3, \hat{N}_B $ ,  and the constant. These 
enter as:

\begin{equation}
  C_3\hat{j}_3 + \omega_0 \hat{N}_B + C_0
\end{equation}
where

\begin{equation}
	C_3 \equiv \ (2\epsilon \cos\theta + 2\Delta^{\prime} \sin\theta +
2\Gamma_2 \xi \sin\theta ) .
\end{equation}

and 
\begin{equation}
C_0 \equiv \omega_0 ( \xi^2  + 1/2) + 2 \Gamma_{1} \xi + \Gamma_{2} \xi.
\end{equation} 
Hence,

\begin{equation}
E_{mn} (\theta,\xi )   =  mC_3 + n\omega_0 + C_0
\end{equation}
and from this we have

\begin{equation}
  \tan\theta  =  -(\Delta^{\prime}/\epsilon)  -  (\xi\Gamma_2)/\epsilon 
\end{equation}
and
\begin{equation}
\xi  =  -\Gamma_1/\omega_0  -  m(\Gamma_2 
\sin\theta)/\omega_0 . 
\end{equation}
As a check of these results we verify that if $\gamma_2$ \ = \ 0 then
$\theta \rightarrow \theta_1$  , and $ \xi \rightarrow \xi_0$   of 
Sections 2 and 3 , with $\hat{H}_{INT} = 0 $.  
\parindent 3em
Corrections to the expressions for $\theta$ and $\xi$ are bilinear or 
quadratic in 
the coupling constants $\Delta' ,\Gamma_{1}$ , and $\Gamma_{2}$, and are
not considered here.

\newpage

{\bf VI. Physical Consequences and Predictions}.

\parindent 3em

{\bf a) Energy and Order Parameters}

\parindent 3em

There are some immediate physical 
consequences of the previous results. First, consider the values of the SC 
and FE order 
parameters in the coherent state $ \vert \Psi_{v} \rangle $ , where the 
variational parameters $ \theta $ and $ \xi $ take on the values just 
determined.

\parindent 3em

For the SC order parameter we  have

\begin{eqnarray}
\eta^{VCS}_{SC} & = &   \langle \Psi_{v} \vert \hat{j}_2 \vert \Psi 
_{v}\rangle \nonumber \\
 & = &   \langle jm \vert \langle n \vert \hat{V} \hat{j}_2 
\hat{V}^{-1} \vert jm \rangle  \vert n \rangle  \nonumber \\
 & = &  \langle jm \vert   (\hat{j}_2 \cos \theta  -\hat{j}_3 \sin 
\theta  ) \vert jm \rangle \nonumber \\
 & = &  \ - m \sin \theta  
\end{eqnarray}

\noindent
or 
\begin{equation}
\eta^{VCS}_{SC}  =  m ( \Delta^{\prime}/\epsilon  \ +  
\xi \Gamma_{2} /\epsilon )/[ 1 + (\Delta^{\prime}/\epsilon  +  \xi 
\Gamma_2/\epsilon)^2 ]^{1/2}.
\end{equation}

\parindent 3em

For the FE order parameter 

\begin{eqnarray}
\eta^{VCS}_{FE}  =   \langle \Psi_{v} \vert (\hat{B}^{\dag}_0 
+\hat{B}_{0}) \vert \Psi_{v} \rangle \nonumber \\ 
 & = & \langle jm \vert \langle n \vert
\hat{V}(\hat{B}^{\dag}_0 + \hat{B}_0) \hat{V}^{-1} \vert jm \rangle \vert 
n \rangle \nonumber \\   
 & = & \langle n \vert  \hat{B}^{\dag}_0 + \hat{B}_0  + 
2\xi \vert n \rangle \nonumber \\  
 & = & 2\xi
\end{eqnarray}

\noindent
or 
\begin{equation}
\eta^{VCS}_{FE}   
=  2( -\Gamma_1/\omega_0 - 
\Gamma_2 ( 1- m \sin \theta ))/\omega_0 .
\end{equation}
As a check we verified that in the absence of $\hat{H}_{INT}$,both order 
parameters $\eta^{VCS}$ revert to their values for the uncoupled systems with
$\Delta^{\prime} \rightarrow \Delta$ and $ \xi \rightarrow \xi_{0}$.

\parindent 3em

Keeping terms up to second degree in $\Delta$ and P we find  the order 
parameters  in the coexisting phase . For the SC order parameter: 

\begin{equation}
\eta^{VCS}_{SC}   =  \eta_{SC} ( 1- \gamma_2 P^2) 
\end{equation}
or
\begin{equation}
\Delta^{VCS} = \Delta^{0}(1-\gamma_{2} P^{2}).
\end{equation}

For the FE order parameter:
\begin{equation}
\eta^{VCS}_{FE}   =  \eta_{FE} (1- \gamma_2 \Delta^2).
\end{equation}
or
\begin{equation}
P^{VCS} = P^{0} ( 1 - \gamma_{2} \vert\Delta\vert^{2}).
\end{equation}

\noindent
In writing the above we identified the order parameters evaluated in
the coexisting state with superscripts VCS, and we used superscipts 0 for the
gap, and polarization values with $\gamma_{2} = 0$, e.g. $\Delta^{0}$ , and 
$P^{0}$ respectively,
and  we expressed the final results using the bare coupling parameter 
$\gamma_{2}$, instead of $\Gamma_{2}$.

\parindent 3em

At this point it is necessary to emphasize again that the sign of the 
parameter 
$\gamma_{2}$ is not fixed by any symmetry argument, but must be determined
from some microscopic considerations and comparison to experiments. As we
pointed out in Section IV above, we have taken 
$\gamma_{2}$ positive in order to implement the competition between
the two types of order. With this sign taken, we immediately conclude that:
the presence of one 
non-zero order parameter will tend to supress the other one .  Thus our 
model supports the "Matthias Conjecture" (1967) \cite{MA6}, which we quote 
here: "Ferroelectricity seems 
to exclude Superconductivity more rigorously than Ferromagnetism seems to 
exclude Superconductivity". We take the conjecture in a weaker sense
for both coexisting superconductivity and ferromagnetism, and for 
superconductivity and ferroelectricity. Namely : in both cases there is an 
"adversarial tendency" , rather than some selection rule prohibiting
coexistence. Since Matthias' statement, numerous  
examples of SC-FM and SC-AFM coexistence/competition have been found
experimentally and studied theoretically. And, as pointed out in the 
Introduction, examples of the SC-FE coexistence are known in 
perovskite-type systems, and in alkalie tungsten
bronze doped systems. Older  work on the $\beta-W$ systems has  considered
superconductivity in "polar metals" ; some examples are in \cite{AN1},
\cite{BI1},\cite{TI1},\cite{BHA1},\cite{BMK1},\cite{BG1}.

\parindent 3em

Returning to our results , in the pseudo-spin model of pure 
superconductivity, the excitation energy at 
each $\vec{k}$ is given by E = $[\epsilon^{2}_{k} + \Delta^{2}_{k}]^{1/2}$.
In our model , $\hat{H}_{DMFA}$ ,  the  coupling will  renormalize 
the gap parameter  $\Delta$ to $\Delta'$.  So as a first approximation we 
can estimate the renormalized SC sector excitation energy and gap order  
parameter as

\begin{eqnarray}
E^{\prime} & = & [\epsilon^{2}_{k} + (\Delta^{\prime }_{k})^{ 2 }]^{1/2} 
; \nonumber \\ 
\Delta^{\prime} & = &  \Delta( 1-\gamma_2 P^{2}).
\end{eqnarray}

\noindent
or
\begin{equation}
E^{\prime} = E - (\gamma_2 \Delta^2 P^2)/ E^2  +  o(\Delta^2 P^4).
\end{equation}

\noindent
The excitation energy $E^{\prime}$ is smaller in the presence of $\vec{P} 
\not=0$, and the "SC gap", $\Delta^{\prime}$ is also smaller than $\Delta$.

\parindent 3em

Similarly we note that the renormalized parameter $\Gamma_1$ causes a 
shift of the minimum potential energy of the FE oscillator from 
$-(\gamma_1 {\cal {E}}^2/\omega_0 )$ to $-(\gamma_1{\cal {E}} - 2\gamma_2 P 
\Delta^2)^2/\omega_0 $, or $ \sim  -(\gamma_1{ E^2}/\omega_0  +
4\gamma_1 \gamma_2  E P \Delta^2/\omega_0 $. Thus there is a smaller 
downward shift of the potential energy of the FE oscillator, and  a less 
stable minimum in the FE sector.

\parindent 3em

Now consider  the energy $E _{mn} ({\theta,\xi})$ for which we have
\begin{equation}
E_{mn} = mC_3 + n\omega_0 + C_0.
\end{equation}
We can identify the SC contribution as $ mC_3$, with $ m = 
(\pm{1/2}, 0 )$ , which includes some FE admixture.  Also  $(n + 1/2) - 
\xi ^2 /\omega_0 $ is the FE oscillator part, including some SC admixture.
Such a  separation is certainly not strict as the renormalized coupling 
constants
$\Delta^\prime , \Gamma_1$ , and $\Gamma_2 $ involve all the interactions.

{\bf b) Experimental Predictions}

\parindent 3em

Some  experimental predictions follow from these results.

\parindent 3em

{\bf i)Pressure Effect on $T_{c}$}

\parindent 3em

From
\begin{equation}
\Delta^{VCS} = \Delta^{0} (1-\gamma_{2}P^{2})
\end{equation}

we can use the BCS result that $ 3.52 k_{B} T_{c} = 2 \Delta  $. We 
identify   $\Delta^{0}$ $\sim 
T^{0}_{c}$ as the "gap" or transition temperature  when there is no FE.

\parindent 3em

According to this result, the SC order parameter in the coexisting state, 
 i.e.  the "gap" $\Delta^{VCS}$  , is decreased from 
$\Delta^{0}$ by non- zero $\hat{P}$ . So,  reducing P 
should increase the gap , and thus our prediction that:  $T_c$ should 
increase as P decreases. In order to test this prediction , we need a 
means of reducing $\hat{P}$ by some applied field.

\parindent 3em

For most perovskite  ferroelectrics, application of (positive) hydrostatic 
pressure $\pi$ will 
decrease P [41] . Thus $dP/d\pi < 0$,where $\pi$ is the applied 
hydrostatic pressure. If, further we assume that the effect 
of pressure on the bare gap (with no FE present) is small,i.e.   
$d\Delta^{0}/d\pi 
\sim 0$, then it is clear that for $\gamma_{2} >0$ and $dP/d\pi <0$ we 
would have as one testable consequence of the above result

\begin{equation}
dT_{c}/d\pi  >  0
\end{equation}

\parindent 3em
Experimentally, the pressure dependence of $T_{c}$ in several
Alkalie Tungsten Bronzes has been measured \cite{BLO1}.
Of particular interest for us are the sodium tungsten bronze family
in  which we distinguished the materials used by Matthias and
others, with $ 0.1 \leq x \leq 1$ , and the newer materials  of Reich et.al. 
with $ x\sim 0.05$.  
For  the "Matthias type " sodium tungsten bronze : $Na_{0.23}W O_{3}$, it 
was found that \begin{equation}
dT_{c}/d\pi \sim  +1.7\times 10^{-5} {^o}K/bar
\end{equation}
It is  tempting to attribute this as confirmation of our prediction 
above. But no direct 
measurement of change of polarization with pressure $\pi$ was made.
We encourage the measurement of the pressure-dependent
ferroelectric polarization $P(\pi)$ in these tungsten bronze systems , which 
will  enable a  test of our prediction.

\parindent 3em

We need also to recall other factors affecting the pressure 
dependence of the gap and $T_{c}$ of superconductors. It is well known 
\cite{VON1} that  a 
pure superconductor will exhibit pressure dependence of $T_{c}$ due to a 
number of factors : a) shift of the Fermi 
level under pressure, thus modifying the density of electron states at 
the Fermi level,b) change of phonon frequency under pressure ,and c) effects 
of pressure on defects , to mention some factors. (Additonal complexity 
is exhibited,for example, in $V_{3}Si$, where hydrostatic pressure and 
uniaxial pressure in the [111] direction give positive coefficients, 
while uniaxial pressure in the [100] direction give a negative 
coefficient.\cite {MW1} ).  It is not simple to separate these 
effects, although for some specific cases theory was developed \cite{TI1},
\cite{BHA1}, \cite{MW1},\cite{BMK1}. Also it is well known that 
even in the very well studied class of $AB_{3}$ A-15, or $\beta-W$ 
compounds ,  the sign of the pressure effect on 
$T_{c}$  can be either positive or  negative for different materials , 
with  "no apparent universality"\cite{VON1} .

\parindent 3em

For the $K$ and $ Rb$
tungsten bronzes the measured [Bloom et.al. \cite{BLO1} ]  sign of the 
slope of pressure-dependence of $T_c$ is
negative; however these  materials have different crystal structures. 
As far as we can determine, there is no report of pressure effect on $T_{c}$ 
for  the materials used by Reich et. al., i.e. sodium tungsten bronzes 
with $x \sim 0.05$.

\parindent 3em

To summarize : the available experiments on the older sodium tungsten bronze
materials of "Matthias type" \cite{BLO1}
agree with our  prediction . If the effect of 
pressure on the "pure" superconducting gap (no FE present) $\Delta^{0}$ is 
small, then the experiments  support the assumption that $\gamma_{2}>0$ 
in our model hamiltonian.

\parindent 3em

{\bf ii)   Magnetic Field Enhancement of  The Ferroelectric Polarization- A 
 Non-Reciprocal  Magneto-Electric Effect}

\parindent 3em
  
Turn now to  the FE order parameter $\eta^{VCS}_{FE}$  , which is 
 the spontaneus FE Polarization. Using:

\begin{equation}
  P^{VCS} = P^{0}(1- \gamma_{2}\vert \Delta \vert^{2}) ,
\end{equation} 
where $P^{0}$ is the "bare" FE polarization in the absence of SC.
For $\gamma_{2} > 0 $ , $P$  is decreased by the presence 
of $\vert \Delta \vert \not= 0$ . Hence ,  the FE polarization should   
increase as $\Delta$ is
decreased. Application of magnetic field will decrease $\Delta$ ,  
ultimately  to zero at $H_c$ , the thermodynamic field. Hence we predict

\begin{equation}
d P/d H > 0 ,H < H_c.
\end{equation}

We are unaware of any experiments testing the predicted increase of the 
ferroelectric polarization  $\hat{P}$ with applied magnetic field for the 
sodium tungsten bronze samples 
of Matthias type $0.1\leq x \leq 1.0$ used either by Matthias et.al. 
[1964,etc.] \cite {MA1}, or by 
Bloom et.al. [1976] \cite {BLO1}. Nor have we found any reports of such 
measurements on 
other SC-FE materials, such as the doped $SrTiO_{3}$,  or the newer sodium 
tungsten bronzes.

\parindent 3em

Our  prediction of a change  of the  spontaneous polarization  with 
applied magnetic field due to 
quenching of the superconductivity is   a prediction of a new 
type of  Magneto-Electric [ME] Effect, or more precisely, a 
Magneto-Polarization effect, since

\begin{equation}
dP/dH \neq 0
\end{equation}
signifies the Magneto-Polarization efffect in the coexistant SC+FE system. 
The usual Magneto-Electric Effect ,
 predicted by Landau-Lifshitz [43], and first discussed for $Cr_{2}O_{3}$ 
by  Dzyaloshinsky,\cite{DZ1}
is a property of materials which are usually ferromagnetic, or 
antiferromagnetic, 
whose total symmetry group includes composite anti-unitary operations 
which will be broken 
such as rotation-reflection (or inversion) combined with time-reversal.
Our coexistent SC-FE system  breaks inversion plus gauge symmetry.
Adding  the applied magnetic field to quench superconductivity 
breaks time reversal $\Theta$ , and places the SC-FE in presence of an 
external $\hat{B}$  field  in a symmetry class for the ME 
effect \cite {DZ2},\cite{FS1}.

\parindent 3em

 Thus a material in the coexistent 
superconducting-ferroelectric state ,in the presence of an applied 
external magnetic field will exhibit broken inversion plus time reversal and
broken gauge  symmetries. Hence our new Magneto-Polarization Effect is 
allowed. If  we now use  one of W. Pauli's famous aphorisms : " 
Anything not  prohibited (by symmetry) is mandatory" we can then predict 
the existence on symmetry  grounds of this new class of Magneto-Electric, or 
Magneto-Polarization Effect  
in  a Superconducting Ferroelectric. There is, however one important remark 
to be made. Unlike  the usual ME effect, in the SC-FE case, the effect is 
not  "reciprocal" between magnetic field and applied electric 
field.

\parindent 3em

 Previous work on the  
Magneto-Electric  Effect  has been reviewed  authoritatively ,\cite{DZ2},
\cite{FS1} .  But, to our knowlege our prediction  is the first 
 of this new Magneto-Electric Effect in a  Superconducting-Ferroelectric 
material.

\newpage

{\bf VII .  Discussion and Conclusions}.

\parindent 3em

The present work was motivated by our  attempt to formulate 
the simplest algebraic model which would embody the relevant dynamical 
symmetries to describe the multicritical point
for coupled superconductivity and ferroelectricity . In 
that framework it is natural to use the pseudo-spin $su(2)$ algebra for 
superconductivity , since this algebra expresses the breaking of gauge 
symmetry embedded in the BCS theory [1957] \cite{BA1}, and
has proven useful in other contexts. We 
introduced the linearly displaced oscillator for the soft-mode phonon
in describing a ferroelectric via the $h(4)$ "Heisenberg Algebra" since 
this  captures several key aspects of the displacive ferroelectric 
transition. Namely the shift to a new equilibrum, which breaks a 
pre-existing 
inversion symmetry and  allows a spontaneous ferroelectric (and 
pyroelectric) moment, and the high dielectric coefficient.The simplest 
invariant coupling of these two order 
parameters which respects both the gauge and the inversion symmetries
requires a biquadratic coupling which then turns to  a bilinear coupling  
when reduced via our "Double Mean 
Field Approximation "[DMFA] . The Hamiltonian $\hat{H}_{DMFA}$ is 
an element in $su(2) \otimes h(4)$.

\parindent 3em

The physical idea motivating the present  work is that the 
 superconducting and ferroelectric transitions  are close to one another 
at a multicritical point. As we know from  Ginzburg-Landau  theory  and  
renormalization group theories of such multicritical behavior, 
 each transition will renormalize the other. A simple 
example of this is that our double mean field approximation (DMFA) 
renormalizes the "bare" coupling constants
(see equation (35)), and leads to shifts in the two order parameters
(see equations (60),(63)) from bare values. Hence  the 
frustration, or reduction, of one order parameter will  enhance the 
other. This is consistent with  the  weaker form of the "Matthias 
Conjecture" on the mutually antagonistic  effects of superconductivity 
and ferroelectricity as shown earlier in  this paper. Matthias  was led 
to this statement in part by his work on the  sodium tungsten bronzes. The 
new work by Reich et.al. on the sodium tungsten  bronzes (in a different 
sodium  composition range)  reopens interest in the particular
questions  related to the superconducting-ferroelectric competition.
We hope that our predictions in this paper  and elsewhere on electrodynamics 
\cite {BIZ} will stimulate further experimental work on these known 
coexistent systems including the titanate and bronze systems.

\parindent 3em

Now we turn to possible relevance of this scenario to the high 
temperature cuprates. As was noted by Weger and collaborators \cite 
{PE1},\cite {PE2},
conventional Eliashberg-McMillan theory of phonon-mediated interaction 
does not consider a multicritical point,nor in particular, the case where 
one phonon is a soft-mode ferroelectric phonon. There is no direct 
neutron scattering evidence for a soft ferroelectric phonon, however there 
is  evidence that a close  near-ferroelectric lattice instability  exists in 
cuprates, and can play a role in the superconductivity mechanism.  

\parindent 3em

The measurements of ionic dielectric  coefficients in several high 
$T_{c}$ cuprates shows exceptionally high values of the dielectric 
coefficients. Thus in YBCO  and LaSrCuO values of $\epsilon \sim 40-50 $ 
were measured below dispersion frequencies of 19 mev and 27 mev, respectively.
These values are  reminiscent of the well-known  near-ferroelectricity 
in the perovskites. [Recall that enhancement of 
SC by nearby  FE was predicted for doped $SrTi0_{3}$ in 1964 by M.L.Cohen 
\cite{MLC1},  and was investigated thoroughly by A.Baratoff  et.al.(1981) 
\cite{BAR1}. By itself doped $SrTiO_3$ is not expected 
to be a superconductor, but rather a nearly ferroelectric 
semiconductor. However,the 
electron-phonon interaction is enhanced by the near-ferroelectricity, and 
possibly  by 
the multivalley conduction band, 
and as a result doped $SrTi0_{3}$ is a superconductor with $T_{c} \sim 
1.5^oK$.]

\parindent 3em

In order to identify the phonons responsible for the high dielectric 
coefficient in the cuprates consider the dispersion of $\epsilon(\omega)$, 
which identifies the c-axis motion of the alkaline earth 
ion (Ba in YBCO, Sr in LaSrCuO) as responsible for the large values of 
$\epsilon$. These are $\bf not$ the phonons inducing the 
superconducting pairing , which are the planar oxygen displacements, at
around 35 mev (transverse) according to \cite{VED1}, and possibly near 70 
mev (longitudinal).

\parindent 3em
We can then suggest the following picture which will relate to our model:
the usual electron-acoustic phonon  and possibly other electron-non-soft 
phonons are responsible for electron-pairing \cite{BAP1}, and their 
effect is present in the "pairing" coefficient 
$\Delta$ in $\hat H_{SC}$ , which carries over to the coupled Hamiltonian
$\hat{H}_{DMFA}$ , which we use in our analysis. The separate 
electron-soft-mode or ferroelectric phonon represented by operators 
$\hat B^{\dag}_{0}$ and $\hat B_{0}$,
with characteristic phonon frequency $\omega_{T0}$ then couples to the 
formed electron pairs, giving the term $\hat H_{INT}$.

\parindent 3em

In the language of Eliashberg-McMillan theory \cite{EL1}, \cite{MCM}  the 
electron-phonon coupling constant $\lambda_{\infty}$ 
for coupling to phonon mode $\Omega_{A}$ is renormalized by proximity to 
a ferroelectric transition, so that the constant becomes frequency-dependent
as $\lambda(\omega) \sim \epsilon^{2}(\omega)\lambda_{\infty}$ \cite{PE1},
and also is large at low frequencies  $\omega \ll \Omega_{A}$ ,i.e. away 
from the physical sheet. This is 
consistent with the small phonon shifts in the superconducting state for 
, e.g. YBCO \cite {ZYH1}  where we can argue that $\lambda(\omega)$ is 
very large. Note , however, large shifts were observed in BaCaHgCuO 
\cite{HAD1} and in organic salts \cite{PIN1}.

\parindent 3em

Making a short digression from the cuprates  
we can remark that the data on $(BEDT-TTF)_{2} Cu(CNS)_{2} $ are illuminating
in this  context.The frequency of the phonon at 2 mev $\bf increases $ by 
almost 
20 $\%$ below $T_c$. The superconducting gap in this material is  2 
$\Delta = 
10 meV$  \cite{WEGX}.  According  to the theory of Zeyher and Zwicknagl 
\cite{ZYH1}
, the phonon frequency should be pushed $\bf down $ below $T_{c}$, while
experimentally it increases. We suggest that the mode at 2 mev is the 
soft or nearly  "ferroelectric" phonon , which is $\bf not $ the one 
responsible  for the pairing ; the phonons responsible for the pairing 
being around 6-8 mev. According to the results  presented in this paper , 
superconductivity suppresses the tendency to ferroelectricity and thus , 
in effect , it "hardens" the relevant phonon.

\parindent 3em

The data on $HgBa_{2}Ca_{3}Cu_{4}O_{10}$ \cite{HAD1} is that the mode at 
30 meV softens at $\ T_{c} $  by about 7 $\%$ , and the mode at 50 meV by 
about 10 $\%$. Thus the direction of the effect is opposite to that in the 
organic superconductor. Hadjiev et.al.\cite{HAD1} account for the effect as 
being due 
to the  enormous anisotropy (which is the cause of the d-wave pairing). 
We do not  consider this in the present work. Also , the frequency of the 
phonon is  considerably higher than that of the "ferroelectric" mode , 
which is 19 mev in YBCO.

\parindent 3em

The cited infra-Red measurements  have shown \cite{KI1},\cite{HE1}, that the 
ionic  dielectric constant is dominated by the c-axis motion of the alkaline 
earth ion . Additional support for the existence of a large ionic dielectric 
coefficient associated with the motion of the Sr ion and its connection 
to the electronic properties of this
material  can be found  from the EXAFS measurements of Pollinger et al
\cite{POLL1} which show an anomaly in the distances of Sr to Oxygen in the 
$LaSrCuO$ material. The large difference in distances between La-O and Sr-O 
supports a large polarization associated with the Sr. The interpretation is in
terms of a Zhang-Rice singlet and an anti-Jahn Teller triplet which are 
nearly degenerate in energy.  Some theoretical calculations of Anisimov 
-Andersen,and Kamimura ,\cite{AND2},\cite{KA1} support the near degeneracy 
,which contributes to the high value of $\epsilon$.

\parindent 3em

Another way to obtain the value of dielectric coefficient is via the small
difference in energies between the Zhang-Rice singlet and the  Jahn-Teller 
triplet. We can call this ${U}_{eff} = U_{bare}/\epsilon$ , which can 
serve as a definition of $\epsilon$. Now, if we assume that this 
$\epsilon$ is
just that ionic dielectric coefficient actually measured in the IR 
experiments , then ${U}_{eff}(\omega)$ will be frequency dependent. 
Further, the small ${U}_{eff}$ is associated
with the lower edge of the "mid-IR" band , and it is described by a 
Jahn-Teller picture, as analysed by Moskvin et.al.(1998) \cite{MOS1}. Added 
support for the identication of a small ${U}_{eff}$ can come from the 
report that there is a softening of a longitudinal mode at a wave-vector 
near the zone boundary. Since this can indicate a tendency for charge 
segregation between the coppers it is consistent with a small ${U}_{eff}$.

\parindent 3em

Returning to the sodium tungsten bronze systems  there is a clear case for 
the  applicability of this model. Pure $WO_{3}$  is ferroelectric , with 
large $\epsilon$, so the reported High $T_{c}$  \cite{RE1} gives a 
picture of a  material  which is a  high $T_{c}$ system in the presence of 
ferroelectricity but without copper which was  previously anticipated to be
ubiquitously associated with  high temperature superconductivity. In the 
new bronzes now under study \cite{RE1}, the superconductivity seems to be  
restricted to portions of the physical surface of the material ; this 
could be  either an inherent effect, or possibly due to a preferential 
sodium  concentration on the surface. The recent STM data of Levi, Millo 
et.al. 
\cite{RE1} indicates  a superconducting gap which is sharp (as for s-wave 
pairing) 
but much smaller than in YBCO , suggesting weak or intermediate coupling 
strength.  Thus if the pairing is due to phonons , the frequency of the 
pairing phonon must be much higher--about 70 meV , as for example due to 
longitudinal oxygen displacements. Ferroelectricity here would involve 
displacements of 
the tungsten atoms, again consistent with two types of phonons playing 
different roles.

\parindent 3em

In short,  the cuprates may present a more complicated situation than the
titanates or bronzes because the role of a latent or near-ferroelectric
soft mode instability  is partly masked -- this is not surprising in view
of the more complicated chemical composition and possible effects of 
magnetic excitations, as well as 
disorder in the cuprates. In the strontium titanate and sodium tungsten 
bronze cases application of the model seems straightforward, and it 
captures symmetry-related aspects of the coexistence-competition of the 
two collective effects. At a microscopic level our picture supports the 
view  that there are two kinds of phonons playing a decisive role in 
these systems with SC-FE coexistence : first,  the usual Frohlich coupled 
acoustic phonons giving the familiar BCS  electron-electron pairing  
effects , and secondly,  the , soft-mode ferroelectric  phonons  which 
yield  the strongly enhanced  static dielectric coefficient (e.g. via the 
Lyddane-Sachs-Teller related physics). As discussed elsewhere \cite{PE1},
\cite{PE2}, the latter phonons give an enhanced Thomas-Fermi screening length ,
which in turn changes the electron-phonon coupling and then the net 
effective electron-electron coupling, so that  when inserted in the 
Eliashberg equations the new physics arises.

\parindent 3em

The dynamical symmetry model proposed here captures certain essential 
features of the superconductor-ferroelectric competition/coexistence. 
Predictions of pressure and magnetic-field effects can be tested and 
will enable the sign of the coupling constant $\gamma_{2}$ to be determined. 
For a given material the same coefficient will regulate the change in $ T_{c}$
under pressure and the ferroelectric polarization under magnetic field
as given in Section VI. Novel electrodynamic effects which are predicted 
for a nearly-ferroelectric superconductor are discussed elsewhere 
\cite{BIZ}.

\parindent 3em

{\bf VIII . Acknowledgements}

\parindent 3em

It is a pleasure to acknowledge helpful discussions and comments at 
various times from Professor Y. Avron ,  Professor C.W.Chu, Professor 
C.S.Ting, Dr.  Mikhail Chernikov, Professor M.E.Fisher, and Professor  
David Schmeltzer. Also help from Ms.Nguyen  Que 
Huong is gratefully acknowledged (J.L.B.).  J.L.B. 
acknowledges with thanks the hospitality of Profesor Ady Mann, and 
Professor Michael Revzen, and the Physics Department, Technion, Haifa, 
Israel,  where part of this work was carried out.

\newpage 

{\bf Appendix 1: Calculation of Variances, and  the DMFA}.

	In order to examine the validity of the DMFA we need to calculate
variances of the operators $\hat j^{2}_{2}$ and $ \hat {Q}^2 $ = 
$(\hat{B}^{\dag}_0 +\hat{B}_0)^2$. First, we estimate the corrections 
to the parameters $\theta$ and $\xi$ due to adding the interaction 
Hamiltonian   $\hat H_{INT}$ , which is characterized 
by the strength $\gamma_2$ , to the non-interacting $\hat h_{SC}$ and
$\hat H_{FE}$. 
	Going back to equation (54) and the renormalized coefficients 
given in equation (55) , after a little algebra , and retaining terms of
lowest (linear ) order in $\gamma_2$ ,  we obtain:

\begin{equation}
\tan \theta  =  (\tan\theta_1) (1 - \gamma_2 P^2)
\end{equation}
and  ,
\begin{equation}
\xi $ = $\xi_0 - 4m \gamma_2 \sin\theta_1 , 
\end{equation} 
where $\sin\theta_1 = (\Delta/\epsilon) /  \sqrt{[ 1 + 
(\Delta/\epsilon)^2]}.$
Next, we note that $\hat j^{2}_2   = (1/4)  ( 2\hat n_k \hat n_{-k} - 2 
\hat j_3 )$ so that we will be able to easily find the expectation values 
directly. We then calculate

\begin{equation}
\hat V_1 \hat j^{2}_2 \hat V^{-1}_1 = 2 \cos^2\theta   \hat j^{2}_2  + 
(1/2)\sin 2\theta \hat j_2 + \sin^2 \theta \hat j^{2}_3 , 
\end {equation}
with $\hat V_1 $  = $exp (i\theta \hat j_1) $  as in equation (39).
We then find

\begin{equation}
\langle(\delta \hat j_2)^2\rangle = 
\langle ( \hat j_2)^2 \rangle  - (\langle \hat j_2 \rangle)^2 
= 3 \cos^2\theta \end{equation}
and then as a figure of merit we can take:

\begin{equation}
\langle (\delta \hat j_2)^2 \rangle / (\langle \hat j_2 \rangle)^2 =
  \cot^2 \theta_1  (1 +(1/2) \gamma^{2}_2 P^2 \sin^2 \theta_1)/ 
(1 - \gamma_{2} P^2).
\end{equation}
Substituting the value of $ \tan \theta_1$ ,  this normalized variance
can then be expressed in terms of the parameters of the Hamiltonian as:

\begin{equation}
\langle  (\delta \hat j_2 )^2 \rangle) /  ( \langle \hat j_2 
\rangle)^2  =  12 (\Delta / \epsilon)^2 ( (1 + \gamma_2 P^2 \sin^2 
\theta_1)/( 1 - \gamma_2 P^2))
\end{equation} 
Apart from the numerical factor , the scale of the variance is
set by $(\Delta / \epsilon)^2 $ which is clearly $ \ll 1 $. This will be 
true as well when the FE Polarization is non zero. Consequently
we verify that the variance of the operator representing SC ( the real part
of the gap operator) is negligable in the variational coherent state.

\parindent 3em

Turning to the variance of $\hat Q $ , we need to calculate it using
the operator $( \hat B^{\dag}_0 + \hat B_0)^2 $, but this is equal to  
$( \hat B^{\dag}_0 )^2 + ( \hat B^2 _0)  + 2 \hat {N}_{B} + 1  $. Then the 
matrix elements are easily evaluated: 

\begin{equation}
\langle n \vert  \hat V_2 \hat Q^2 \hat V^{-1}_2 \vert n  \rangle
=  2n + 4 ( \xi )^2  + 1.
\end{equation}
and
\begin{equation}
 \langle n \vert \hat V_2 \hat Q \hat V^{-1}_2 \vert n \rangle  = 2 \xi.
\end{equation}
The normalized variance is then:

\begin{equation}
( \langle \hat Q^2 \rangle  - (\langle \hat Q \rangle)^2 ) / 
(\langle \hat Q \rangle)^2   = ( (2n+1)/4(\xi)^2).
\end{equation} 
Since $\xi  \sim P/2  + o( \gamma_2) $ , and  P will be large in the FE 
state , we verify that $ ( \langle \delta \hat Q^2 \rangle ) \ll 1 $
in the variational coherent  state $ \vert \Psi_v \rangle $.  

\parindent 3em
	We may then conclude that the DMFA , in the variational coherent 
state approximation with wave function $\vert \Psi_v \rangle $ is a 
self-consistent approximation.

\bigskip

\newpage

{\bf Appendix 2: Correlators}

\parindent 3em

Using our model we can  examine the thermal average of two-time 
correlators of 
the basic operators of the model: the FE phonon displacement operator 
$\hat{Q}$, and the real part of the pair operator $\hat{j}_{2}$. 
The thermal average of the  two time correlator of an operator 
$\hat{O}_{\alpha}$ is:

\begin{equation}
\langle \langle \hat{O}_{\alpha}(t) \hat{O}_{\alpha}(0) \rangle \rangle =
(1/Z) Tr (exp 
-(\beta\hat{H}))exp(i\hat{H}t)\hat{O}_{\alpha}exp(-i\hat{H}t)O_{\alpha}
\end{equation}
The partition function is  $Z = Tr e^{-\beta\hat{H}}$ with $\beta 
= (k_{B}T)^{-1}$. Cross correlators can be similarly defined, by 
replacing one
of the  $\alpha$ subscipts by a different index referring to another 
of the operators.

\parindent 3em
Calculation of these correlators is not possible if we use the full 
Hamiltonian, or even if we use $\hat{H}_{DMFA}$. However if we denote
the  transformed $\hat{H}_{DMFA}$ by $\hat{H'}$

\begin{equation}
\hat{H'} = \hat{V} \hat{H}_{DMFA} \hat{V}^{-1}
\end{equation}
and evaluate $\hat{H'}$ at $\theta = \theta_{VCS}$ and $\xi = \xi_{VCS}$, 
we may then take as a good approximation
\begin{equation}
\hat{H'} \sim C_{3} + \omega_{0}\hat{N}_{B} + C_{0}
\end{equation}

Making this Ansatz will permit us to easily evaluate the operators in 
Heisenberg picture,and then evaluate the traces. This is
consistent with retaining the dominant leading order in the coupling 
parameters $\Delta$,$\gamma_{1}$ , and $ \gamma_{2}$.

\parindent 3em

We first  work out the displacement-displacement correlator. Using invariance
of the trace we can then write

\begin{equation}
\langle \langle \hat{Q}(t)\hat{Q}(0) \rangle \rangle  = (1/Z) 
Tr[\hat{V}e^{-\beta\hat{H}}\hat{V}^{-1}\hat{V}e^{i\hat{H}t}\hat{V}^{-1}
\hat{V}\hat{Q}\hat{V}^{-1}
\hat{V}e^{-i\hat{H}t}\hat{V}^{-1}\hat{V}\hat{Q}\hat{V}^{-1}]
\end{equation}

\parindent 3em
Using the properties of the operators $\hat{Q}$ and $\hat{H}$ transformed 
under $\hat{V}$ we obtain an intermediate result

\begin{equation}
\langle \langle \hat{Q}(t) \hat{Q}(0) \rangle \rangle = (1/Z) Tr 
[e^{i\beta\omega_{0}\hat{N}_{B}}(e^{i\omega_{0}t}\hat{B}^{\dag}_{0}
+e^{-i\omega_{0}t}\hat{B}_{0} + 2\xi) + 2\xi e^{-\beta 
\omega_{0}\hat{N}_{B}}(\hat{Q} + 2\xi)].
\end{equation}

\parindent 3em
As only the lattice oscillator Bose  operators have survived we now
carry out the diagonal summation over the harmonic oscillator 
quantum number $n$. We then find

\begin{equation}
Z(\beta) = e^{\beta \omega_{0}/2} \times [2 \sinh(\beta\omega_{0})]^{-1}
\end{equation}

and finally
\begin{equation}
\langle \langle \hat{Q}(t) \hat{Q}(0) \rangle \rangle = 4\xi^{2}+
e^{-i\omega_{0}t} + 2Z(\beta) \cos \omega_{0}t.
\end{equation}

\parindent 3em

The same correlator can be evaluated for a Hamiltonian $\hat{H}^{0}_{FE}= 
\omega_{0}
(\hat{N}_{B} + 1/2) $ ,without the linear displaced term ; it gives the 
same result as above without the term $4 \xi^{2}$. The value of $\xi$ to 

be taken here is $\xi_{VCS}$ from the solution of the variational
problem.

\parindent 3em

In exactly the same way , but now doing the trace over the 4 electronic 
states with $ m = 0, 0, \pm 1/2$ we obtain:

\begin{equation} 
\langle \langle \hat{j}_{2}(t)\hat{j}_{2}(0) \rangle \rangle  = 
(\sin^{2}\theta_{VCS}\times \cosh \beta C_{3})[4(1 + \cosh\beta C_{3}]^{-1}.
\end{equation}
\newpage
{\bf Appendix 3-Some $su(2)$ results}.

 In this Appendix we record some known results, which will be useful in 
various calculations throughout the paper.If $\hat{a}_{k\uparrow},\hat 
{a}_{k\uparrow}^{\dag}$ are the electron
annihilation/creation operators for wave vector $\vec{k}$, spin $(\uparrow)$,
the relevent pair operators are defined by

\begin{equation}
\hat{b}_{k}^{\dag}  \equiv  \hat{a}_{k\uparrow}^{\dag}
\hat{a}_{-k\downarrow}^{\dag}  ; \ \hat{b}_k \  =  \
(\hat{b}^{\dag}_k)^{\dag} ; \ \
 \  \hat{n}_k \  \equiv \ \hat{a}_{k\uparrow}^{\dag}
\hat{a}_{k\uparrow}.
\end{equation}

\noindent
The $su(2)$ pseudospin algebra  at each $\vec{k}$ is generated by

\begin{equation}
\hat{j}_{1k} \equiv (-i/2)(\hat{b}_{k}^{\dag}-\hat{b}_{k}),
\end{equation}

\noindent
\begin{equation}
\hat{j}_{2k} \equiv (1/2)(\hat{b}_{k}^{\dag}+\hat{b}_{k}),
\end{equation}

\noindent
\begin{equation}
\hat{j}_{3k} \equiv (-1/2)(\hat{n}_{k}+\hat{n}_{-k} -1)
\end{equation}
so that
\begin{equation}
[\hat{j}_{pk},\hat{j}_{qk}]_- \ = \ i \epsilon_{pqr}\hat{j}_{rk}, \ {\rm 
where}
   \ \ (p,q,r) = (1,2,3)
\end{equation}

 A basic set of states for each sector $\hat{h}_{k}$ can be
obtained by starting from $su(2)$ eigenstates labelled  by a pair of
indices . Thus, since the eigenvalues of $\hat{n}_{{\pm}k}$ are
$(0,1)$ the states can be labelled by the eigenvalues $(n_{k},n_{-k})$. Or,
we can use the $(j m)$ labels of the ket $\vert jm \rangle$ referring to
$su(2)$  (suppressing $k$), which is an eigenstate of $\hat{j}^{2}$ and
of $\hat{j}_{3}$. Here

\begin{equation}
\hat{j}_{3} \vert jm\rangle  \ = \ m\vert jm\rangle ; \ \
\hat{j}^{2}\vert jm\rangle  \ = \ j(j+1)\vert jm\rangle.
\end{equation}

\noindent
where we note that $\hat{j}^2$ \ = \ $(\hat{n}_{k}\hat{n}_{-k}
+\hat{j}_{3k}+\hat{j}^{2}_{3k})$. Thus $\hat{j}^{2}$ ,  as well as
$\hat{j}_{3k}$ , depends only on the number operators for ${\pm{k}}$
and so their eigenvalues are easily computed from the allowed
eigenvalues $(n_{k} ,n_{-k})$.  Also the same states are
obtained by applying creation
operators to the vacuum state like $\hat{a}^{\dag}_{k}\vert 0\rangle$
etc. Enumerating these states at each k we have:

\begin{equation}
(0,0) \sim \vert 3/4, 1/2  \rangle  \sim \vert 0 \rangle
\end{equation}

\begin{equation}
(0,1) \sim \vert  0, 0   \rangle  \sim \hat{a}^{\dag}_{-k \downarrow}
\vert 0 \rangle
\end{equation}

\begin{equation}
(1,0) \sim \vert 0 , 0 \rangle  \sim \hat{a}^{\dag}_{k \uparrow} \vert 0
\rangle
\end{equation}

\begin{equation}
(1,1) \sim \vert 3/4 ,-1/2 \rangle \sim \hat{a}^{\dag}_{k\uparrow}
\hat{a}^{\dag}_{-k \downarrow} \vert 0 \rangle  \sim \hat{b}^{\dag}_{k}
\vert 0 \rangle
\end{equation}

In each line, the states are presented in order, labelled by  $( n{_k},
n{_{-k}} )$ , by $\vert j,m \rangle $ , and finally  by   creation
operators applied to the vacuum $ \vert 0 \rangle $ . In constructing  
the eigenstates for $ \hat h _{k}$  these  basic states of the 
free hamiltonian will be rotated to produce the $su(2)$-coherent states, 
which are eigenstates of $\hat{h_{k}}$. Recall \cite{BI4} that the states
$(0,0)$ and $(1,1)$ are the basis for the irreducible representation 
$D^{(1/2)}$, and $(0,1)$ and $(1,0)$ are bases for $D^{(0)}$ , of 
$su(2)$.(N.B. The latter two states are a time reverse pair for the
corepresentation of $su(2) \otimes \Theta$).

\parindent 3em
The eigenstates of $\hat{j}_{3k}$ are the basic set $\vert jm\rangle$
given above. To obtain the eigenstates
$\vert\psi_{k}\rangle$  or $\vert\Psi_{ord} \rangle$ , of the original
problem

\begin{equation}
\hat{h}_k\vert\psi_k\rangle \ = \ \lambda_k\vert\psi_k\rangle
\end{equation}

\noindent
transform by $\hat{U}_1$ as follows

\begin{equation}
\hat{U}_{1}\hat{h}_{k}\hat{U}^{-1}_{1}\hat{U}_{1}\vert \psi_{k}\rangle \
= \ \lambda_{k} \hat{U}_{1}\vert \psi_{k}\rangle
\end{equation}

\noindent
and then we find

\begin{equation}
\lambda_k  = \ m E_{k}, \ {\rm and} \  \vert\psi_{k}\rangle \ = \
\hat{U}^{-1}_1\vert jm\rangle.
\end{equation}

\noindent
Now we can evaluate the order parameter in the state $\vert \psi_{k}
\rangle$ = $\vert \Psi_{ord} \rangle$:

\begin{eqnarray}
\eta_{SC} & \equiv &
\langle\psi_{k}\vert \hat{j}_{2k}\vert\psi_{k}\rangle  
  =   \langle jm\vert\hat{U}_{1}\hat{j}_{2k}\hat{U}^{-1}_{1}\vert
jm\rangle = \langle jm\vert(\hat{j}_{2k} \cos\theta_{1} + \hat{j}_{3k}
\sin\theta_{1})\vert jm\rangle 
   =  m \sin\theta_{1}
\end{eqnarray}

\noindent
or
\begin{equation}
\eta_{SC}  =  m \Delta_k/\sqrt{\Delta^{2}_{k} +
\epsilon^{2}_{k} }.
\end{equation}

\noindent
The lowest (ground state) energy $\lambda_{G}$ occurs when
$ m \ = \ -1/2$, and the corresponding order parameter $\vert\eta_{SC}\vert
 \not = 0 $, and $ \lambda_{G}  =  (-1/2) E_{k} $. The ground eigenstate
$\vert \psi_{kG} \rangle $ is
the BCS pair state which can be written in several alternate forms. These
are

\begin{eqnarray}
 \vert\psi_{kG}\rangle & = &  exp (-i\theta_{1}\hat{j}_{1k})\vert
3/4,-1/2\rangle\\
 & = & exp -[(\theta_{1}/2)(\hat{b}^{\dag}_{k}-\hat{b}_{k})]
\hat{b}^{\dag}_{k}\vert0\rangle \\
 & = &  \hat{U}^{-1}\vert 3/4,-1/2\rangle.
\end{eqnarray}

or
\begin{equation}
\vert \psi_{kG} \rangle  =  -( u_{k} + v _{k}\hat{b}^{\dag}_{k})\vert0\rangle
\end{equation}

\noindent
with $u_{k}   \equiv   -sin(\theta_{1}/2)$ ;  $ v_{k}   \equiv
 cos(\theta_{1}/2)$.

\parindent 3em

In passing we note that we can recover the BCS gap equation from the above.
The  $\eta_{SC}$ given above is for one sector k. If we assume that
$\Delta_{k} = V\Delta $ ,i.e. is independent of $k$ and that the "global" 
order parameter is \begin{equation}
\Delta = \sum_{k}\eta_{SC(k)}
\end{equation}

we may then  obtain  a self-consistent equation for the total gap
function as:

\begin{equation}
\Delta =  \sum_{k} \eta_{SC(k)} =  (1/2)\sum_{k}(V \Delta_{k})/
\sqrt{\epsilon^{2}_{k} + \Delta^{2}_{k} }
\end{equation}

and substituting the above we have:

\begin{equation}
2 =  \sum_{k} ( V)/ \sqrt{\epsilon^{2}_{k} + \Delta^{2}_{k}}
= V \int \rho({\epsilon}) d{\epsilon}  / \sqrt{ \epsilon^{2} + \Delta^{2}}
\end{equation}
Here,we changed variables to energy , and $ \rho (\epsilon) $ is the density
of states (at the Fermi level) This  reproduces the familiar result

\begin{equation}
(2/V) = \int \rho({\epsilon}) d{\epsilon} / \sqrt{\epsilon^{2}
+ \Delta^{2}}.
\end{equation}

.

\end{document}